\documentclass[a4paper,12pt]{article}

\usepackage{color}
\usepackage[english]{babel}
\usepackage{amssymb,amsmath}
\usepackage{graphicx}
\usepackage{hyperref}

\newcommand{\be}{\begin{equation}}
\newcommand{\ee}{\end{equation}}
\newcommand{\beqa}{\begin{eqnarray}}
\newcommand{\eeqa}{\end{eqnarray}}

\newcommand\m{\mu}

\newcommand\n{\nu}

\renewcommand\l{\lambda}

\begin{document}
\renewcommand\bibname{\Large References}

\title{\vspace{-2cm} 
\begin{flushright}
{\normalsize
CERN-PH-TH-2015-043} \\
\vspace{-0.5cm}
{\normalsize INR-TH/2015-008}
\end{flushright}
\vspace{0.5cm} 
{\bf On stability of electroweak vacuum\\ during inflation}}

\author{
A. Shkerin$^{1,2}$\thanks{{\bf e-mail}: andrey.shkerin@epfl.ch}~,~ 
S. Sibiryakov$^{1,2,3}$
\\[2mm]
{\normalsize\it $^1$Institut de Th\'eorie des Ph\'enom\`enes Physiques, EPFL, CH-1015 Lausanne, Switzerland}\\[1.5mm]
{\normalsize\it $^2$ Institute for Nuclear Research of the
Russian Academy of Sciences,}\\[-0.5mm]
{\normalsize\it 60th October Anniversary Prospect, 7a, 117312
Moscow, Russia}\\[1.5mm]
{\normalsize\it  $^3$ CERN Theory Division, 
CH-1211 Geneva 23, Switzerland}
}
\date{}

\maketitle

\begin{abstract}
We study 
Coleman -- De Luccia tunneling of the Standard Model Higgs field
during inflation in the case when the electroweak vacuum is
metastable. We verify that the tunneling rate is exponentially suppressed.
The main contribution to the suppression is the same as
in flat space-time.
We analytically estimate the
corrections due to the expansion of the universe and an effective
mass term in the Higgs potential
that can be present at inflation.
\end{abstract}

\section{Introduction}

\begin{figure}[t]
\centerline{\vspace{-0.5cm}
\includegraphics[width=10cm]{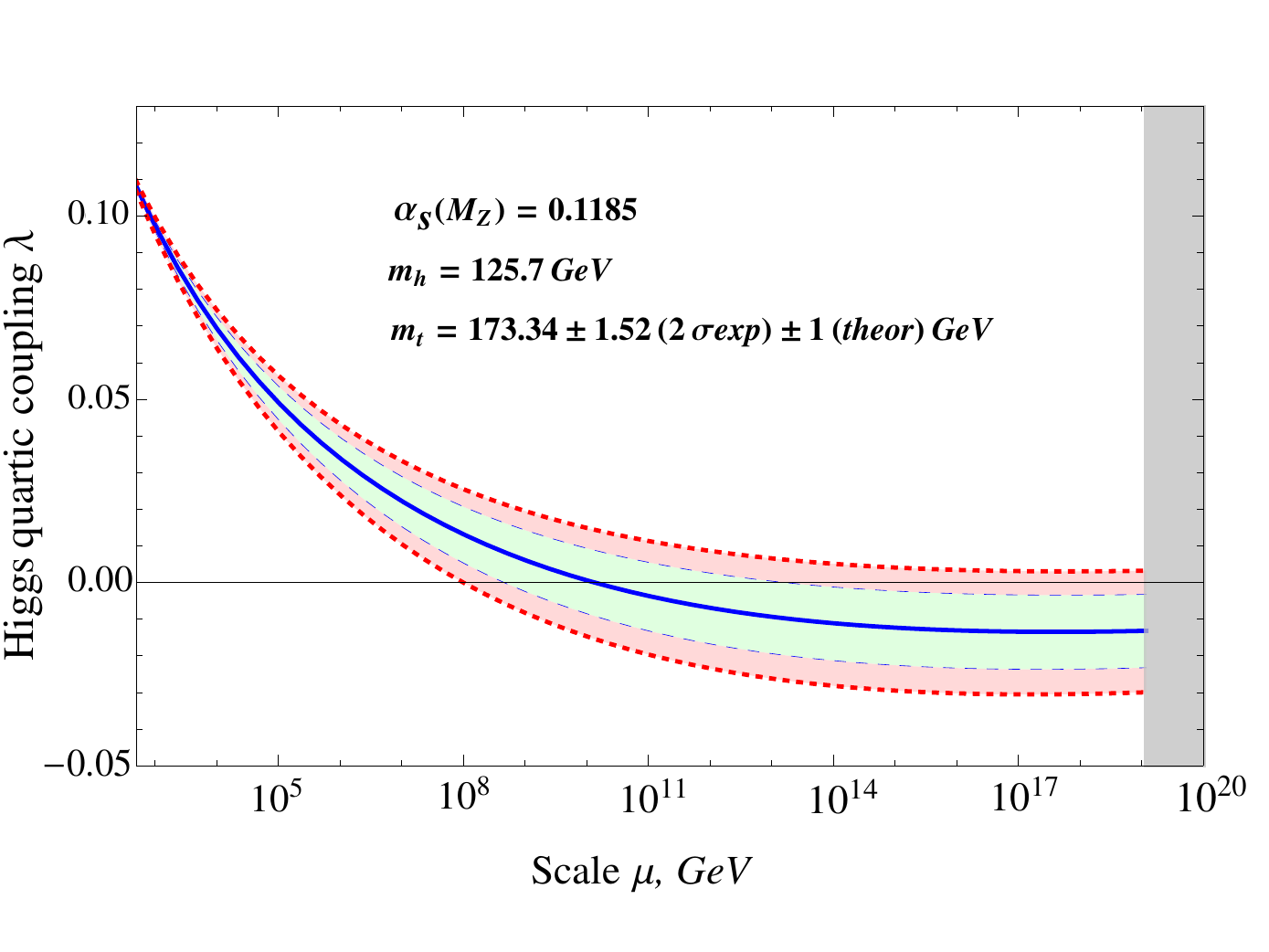}}
\caption{Running of the Higgs quartic coupling in the Standard
  Model at NNLO in the $\overline{\mathrm{MS}}$ scheme. The RG equations
  are solved 
using the code available at \cite{code} based on 
\cite{Chetyrkin:2012rz,Bezrukov:2012sa}. Blue solid line
corresponds to the best-fit values of the Standard Model parameters
\cite{PDG}. Blue dashed lines correspond to  
$2\sigma$ experimental uncertainty in the measurement of the top-quark
mass~\cite{ATLAS:2014wva} and red dotted lines --- to the
theoretical uncertainty discussed in \cite{Bezrukov:2014ina}.
The plot is restricted to the scales smaller than the Planck mass
$M_p=1.22\cdot 10^{19}$~GeV.
\label{fig:1}}
\end{figure}

\begin{figure}[t]
\centerline{\vspace{-0.5cm}
\includegraphics[width=8.5cm]{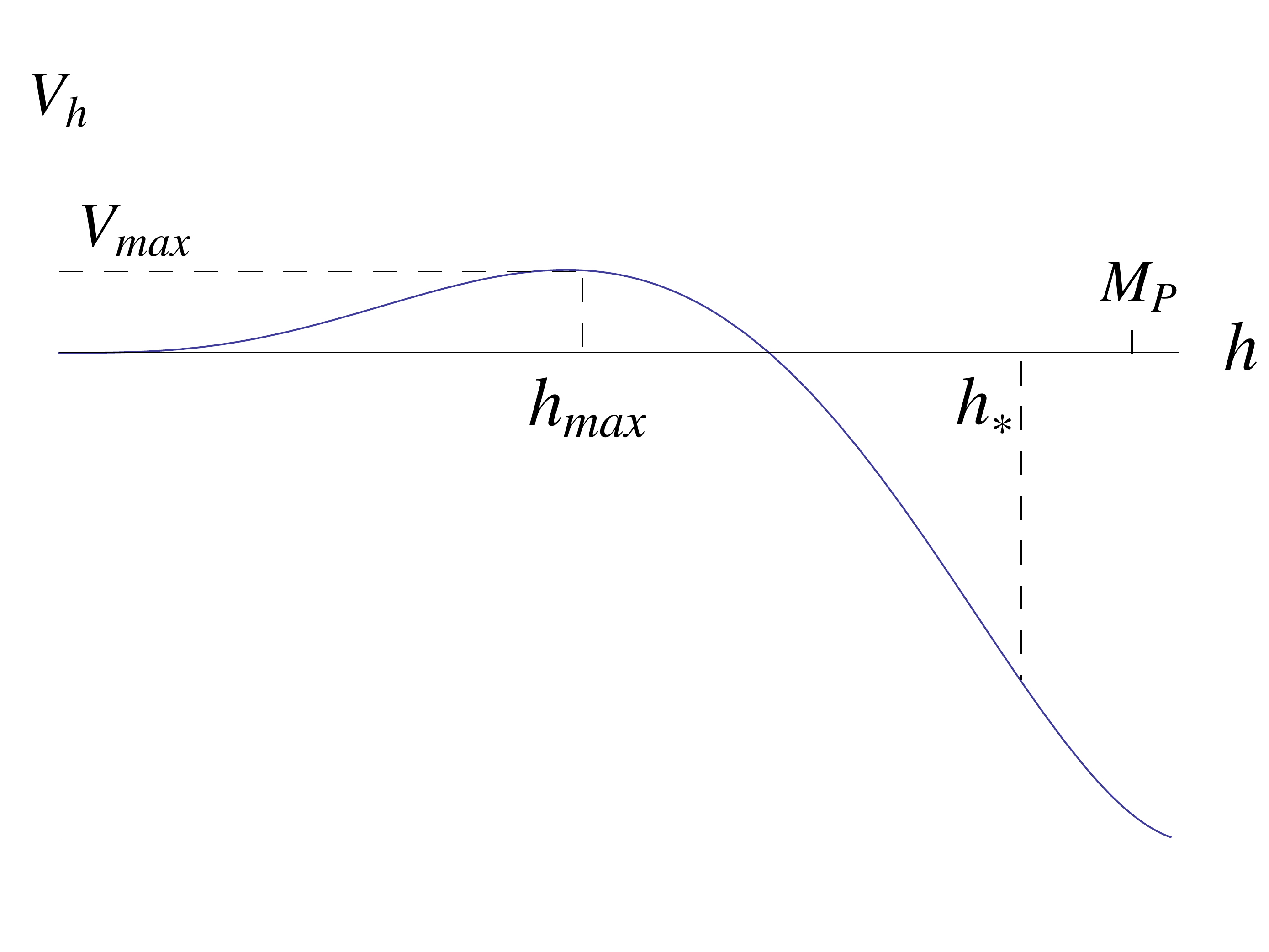}
\vspace{-0.4cm}}
\caption{
Schematic form of the effective Higgs potential (not to scale).\label{fig:2}}
\end{figure}

At tree level the Standard Model (SM) Higgs potential has an absolute
minimum corresponding to the electroweak (EW) vacuum. The loop
corrections change the picture drastically.
They modify the effective potential for the Higgs field through the
renormalization group (RG)
running of the Higgs quartic coupling $\l$ 
\cite{Bezrukov:2012sa,Degrassi:2012ry}. 
The precise evolution of
$\l$ strongly depends on the values of the Higgs and top-quark
masses. It is still possible, within uncertainties of the top mass, 
that $\l$ stays positive all the way up to the
Planck scale~\cite{Bezrukov:2014ina}. However, for the current 
best-fit values of the SM parameters, $\l$ changes sign at large
RG scale $\m_0\sim 10^{10}~\mathrm{GeV}$ and reaches a negative
minimum at
$\m_*\sim 10^{16}\div 10^{18}~\mathrm{GeV}$,
see Fig.~\ref{fig:1}. 
It is worth stressing that this RG evolution is obtained under the
assumption of no new physics interfering with the running of $\l$.
As a result, the effective Higgs potential\footnote{We neglect the SM
  mass term which is tiny compared to all contributions appearing below.}
\be
\label{Higgspot}
V_h=\frac{\l(h)h^4}{4}
\ee
goes much below the EW vacuum at large values of the field, as shown
schematically in  
Fig.~\ref{fig:2}. This makes the EW vacuum metastable.

While in a low density, low temperature environment  
characteristic of the present--day universe the SM vacuum is
safely long-lived    
\cite{Degrassi:2012ry}, the situation
may be
different during primordial inflation. Indeed, most inflationary
models predict the Hubble expansion rate during inflation $H_{inf}$ to
be much higher than the measured Higgs mass. Thus, if the Higgs 
does not have any other couplings besides those present in SM, it
behaves at inflation as an
essentially massless field and develops fluctuations of order
$H_{inf}$. Denote by $h_{max}$ the value of $h$
corresponding
to the top of the barrier separating the EW vacuum from the run-away
region. 
Then,
even if $h$ is originally placed close to the origin, it will roll
beyond the barrier with order-one probability for $H_{inf}>h_{max}$
\cite{Espinosa:2007qp,Lebedev:2012sy,Kobakhidze:2013tn,
Fairbairn:2014zia,Enqvist:2014bua,Hook:2014uia}.     

A simple cure to the problem is to endow the Higgs 
with an effective mass $m_{eff}\gtrsim H_{inf}$ during the inflationary
stage. This can be due, for example, to a 
non-minimal coupling to gravity\footnote{We work in the
  signature $(-,+,+,+)$, so that the curvature of de Sitter space is
  positive, $R=12H^2_{inf}$.}, $V_{hR}=\xi R h^2/2$
\cite{Espinosa:2007qp,Herranen:2014cua}, or 
a coupling between
$h$ and the inflaton field\footnote{We assume that the inflaton is
  distinct from the Higgs, unlike the case of Higgs inflation
  \cite{Bezrukov:2007ep}.} $\phi$ of the form 
$V_{h\phi}=f(\phi)h^2/2$ \cite{Lebedev:2012sy,Fairbairn:2014zia}.
This raises the potential barrier and suppresses the over-barrier
transitions. In this situation the EW vacuum is still able to decay via
quantum tunneling. 

Tunneling from a false vacuum in (quasi-) de Sitter space-time
can proceed in two distinct regimes: via the Hawking--Moss (HM) instanton 
\cite{Hawking:1981fz} which describes quantum jumps on top
of the potential barrier, or via Coleman--De Luccia (CDL) bounce
\cite{Coleman:1980aw} corresponding to genuinely under-barrier
penetration. While HM transitions have been extensively discussed in
connection with the Higgs behavior during inflation (see e.g. 
\cite{Espinosa:2007qp,Kobakhidze:2013tn,Enqvist:2014bua,Hook:2014uia}), 
the CDL tunneling
is usually discarded with the common lore that it is sufficiently
suppressed. However, to the best of our knowledge, a
verification of this assertion is missing in the literature\footnote{Note that
  the thin-wall approximation, which is often invoked in the analysis
  of the CDL tunneling and which makes the
  exponential suppression manifest, is not applicable in the case
  of the Higgs field.}. Moreover, Ref.~\cite{Kobakhidze:2013tn} which
explicitly addressed this question has reported an opposite
result that the CDL decay of the EW vacuum is enhanced, instead of being
exponentially suppressed. If true, this would pose a serious challenge
for the stability of the EW vacuum during inflation.  

The purpose of this {\it letter} is to clarify the above issue. We
will estimate the CDL tunneling rate and confirm that it is
exponentially suppressed. The suppression exponent will be found to be
essentially the same as in flat space-time, up to small corrections which we
will estimate analytically.

\section{Bounces in de Sitter space}

In this section we assume that the energy density of the universe is
dominated by the inflaton with negligible back-reaction of the Higgs
field on the metric. The validity of this assumption will be discussed
later. Then, neglecting the slow-roll corrections, we arrive to the
problem of a false vacuum decay in external de Sitter space-time. This
process is described by the Euclidean version of the Higgs action  
\begin{equation}
\label{Eucact}
S_E=\int d^4x\sqrt{g_E}\bigg(\frac{1}{2}g_E^{\m\n}\partial_{\mu}h\partial_{\nu}h+V_{h}(h)\bigg)\;,
\end{equation}
where $g_{E\,\m\n}$ is the metric of a 4-dimensional sphere, which is
the analytic continuation of the de Sitter
metric \cite{Coleman:1980aw} (see also \cite{Rubakov:1999ir}), 
\begin{equation}
\label{dSmetric}
ds_E^2=d\chi^2+\rho^2(\chi)d\Omega_3^2~,~~~~
\rho=\dfrac{1}{H_{inf}}\sin(H_{inf}\chi)~,~~~
0\leq\chi\leq \frac{\pi}{H_{inf}}\;.
\end{equation}
Here $d\Omega_3$ is the line element on a unit 3-sphere. We search for
a smooth solution of the Higgs equations of motion following from
(\ref{Eucact}). Assuming $O(4)$ symmetry, one reduces the
action to  
\begin{equation}
\label{mainaction}
S_E=2\pi^2\int_0^{\pi/H_{inf}}
\!\!\! d\chi\,\rho^3\left(\frac{h'^2}{2}+V_{h}\right)\;,
\end{equation}
which yields the equation for the bounce $h_b(\chi)$,
\begin{subequations}
\label{meqs}
\begin{equation}
\label{mainequation}
h_b''+3H_{inf}\,\mathrm{ctg}(H_{inf}\chi)\,h'_b=\frac{dV_{h}}{dh}\;.
\end{equation}
To be regular, the solution must obey the boundary conditions,  
\be
\label{bcs}
h_b'(0)=h_b'(\pi/H_{inf})=0\;.
\ee
\end{subequations}
The
probability of false vacuum decay per unit time per unit
volume scales as
\be
\label{HMprob}
\frac{dP}{dt d{\cal V}}\propto \exp(-S_E)\;,
\ee
where the action is evaluated on the solution $h_b(\chi)$.

\paragraph{Hawking--Moss instanton.}
Equations (\ref{meqs}) always have a constant solution with the Higgs
field sitting on top of the potential barrier, $h_b=h_{max}$ (see
Fig.~\ref{fig:2}).  
This instanton can be interpreted as describing
the over-barrier jumps of the Higgs field due to non-zero de Sitter
temperature, $T_{dS}=H_{inf}/(2\pi)$ \cite{Gibbons:1977mu}. 
The rate of such transitions is given by (\ref{HMprob}) with the
action
\be
\label{HMact}
S^{(HM)}_E=\frac{8\pi^2}{3}\frac{V_{max}}{H_{inf}^4}\;.
\ee
The transition rate is exponentially suppressed if $H_{inf}\lesssim
V_{max}^{1/4}$. In the pure SM $V_{max}^{1/4}$ is of order $10^9$~GeV
\cite{Degrassi:2012ry} implying that the EW vacuum is stable with
respect to HM transitions whenever $H_{inf}<10^9$~GeV and unstable
otherwise. In the latter case new contributions into the Higgs
potential that raise $V_{max}$ are required to stabilize the SM
vacuum. A simple option is to endow
$h$ with an effective mass $m_{eff}$ during inflation. The potential becomes
\be
\label{effpot}
V_{h}=\frac{\l(h)\,h^4}{4}+\frac{m_{eff}^2h^2}{2}\;.
\ee
For $H_{inf}\gtrsim 10^{10}$~GeV the qualitative picture is captured
by neglecting the slow logarithmic dependence of the coupling on the
field and normalizing it at a fixed scale above $\m_0$, so that $\l$
is negative and is of order $0.01$ in the absolute value. 
This gives for the position and height of the potential barrier,
\be
h_{max}=\frac{m_{eff}}{\sqrt{|\l|}}~,~~~
V_{max}=\frac{m_{eff}^4}{4|\l|}
\ee 
leading to the instanton action,
\be
\label{HMact1}
S^{(HM)}_E=\frac{8\pi^2}{3|\l |}\bigg(\frac{m_{eff}}{H_{inf}}\bigg)^4\;.
\ee
As expected, the transitions are strongly suppressed provided the mass
is bigger than $|\l|^{1/4}H_{inf}$. Note that for these values of
the mass $h_{max}$ lies above $\m_0$, which justifies our
approximation of constant negative $\l$. For the case when the Higgs
mass is due to non-minimal coupling to gravity one has
$m_{eff}^2=12\xi H_{inf}^2$, so that the suppression (\ref{HMact1})
does not depend on the Hubble parameter and is large already for
$\xi\gtrsim 0.1$  
\cite{Espinosa:2007qp,Herranen:2014cua,Kamada:2014ufa}.

\paragraph{Coleman--De Luccia bounce.}
Another decay channel is described by inhomogeneous solutions of
(\ref{meqs}) which interpolate between the false vacuum and a value
$h_*$ in the run-away region. These
correspond to genuinely under-barrier tunneling. 
To understand their properties, let us first neglect the
running of $\l$ normalizing it at a high enough scale, so
that $\l<0$. 
If we further neglect the mass and space-time curvature,
we obtain the setup of tunneling from the top of an inverted quartic potential
in flat space. This is described by a family of bounces,  
\begin{equation}
\label{Lipaton}
h_{\bar\chi}(\chi)=\sqrt{\frac{8}{|\lambda|}}\,\frac{\bar\chi}{\chi^2+\bar\chi^2}\;,
\end{equation}
parameterized by their size $\bar\chi$. The action of these solutions is
independent of $\bar\chi$ due to the classical scale invariance of the setup,
\begin{equation}
\label{LipatonAction}
S_E=\frac{8\pi^2}{3|\lambda|}\;.
\end{equation}
The mass and finite Hubble rate break the degeneracy. Assuming that
the size of the instanton is small compared to the length
\be
\label{scale}
l=\min(m^{-1}_{eff},H^{-1}_{inf})
\ee
characterizing the breaking of scale invariance, one can estimate the
corrections to the bounce action perturbatively. Substituting
(\ref{Lipaton}) into (\ref{mainaction}) and expanding to the order
$O((l/\bar\chi)^2)$ we obtain, 
\begin{equation}
\label{mHcorrections}
S_E^{(CDL)}(\bar\chi)=\frac{8\pi^2}{3|\l|}
\big[1+3(m_{eff}^2-2H_{inf}^2)\bar\chi^2\log(l/\bar\chi)\Big]\;,
\end{equation}
where we have kept only the log-enhanced contributions. The tunneling
rate is given by the configuration minimizing the action. If
$m_{eff}^2>2H_{inf}^2$
the minimal suppression is reached at the configuration
of zero size\footnote{A proper interpretation of this singular
  bounce
is given within the formalism of constrained
  instantons~\cite{Affleck:1980mp}.}, $\bar\chi=0$, and coincides with
the flat-space 
result (\ref{LipatonAction}). 
One observes that in this case the assumption $\bar\chi\ll l$ is 
justified.
In the opposite case, $m_{eff}^2<2H_{inf}^2$,
the correction due to the expansion of the universe dominates and
makes the solution spread over the whole 4-sphere. We have
checked numerically that the only solution in this case is the HM
instanton. 

We now restore the running of couplings which provides additional
source of the scale invariance breaking.
This
enters into the calculations through the loop corrections in the
instanton background. For instantons of the size smaller than $l$
these corrections can be evaluated neglecting
both the mass $m_{eff}$ and the Hubble $H_{inf}$. Thus, they are the
same as in the flat space \cite{Isidori:2001bm} and roughly amount to
substituting in (\ref{mHcorrections}) the coupling constant evaluated
at the scale of inverse instanton size, $\m=\bar\chi^{-1}$. Numerically,
for the best-fit values of the SM parameters, this dependence on
$\bar\chi$ turns out to be much stronger than the one introduced by the
effective mass and the Hubble expansion. This freezes the size of the
instanton at the value corresponding to the minimum of the running
coupling constant, $\bar\chi^{-1}_*\approx \m_*\sim 10^{16}\div
10^{18}$~GeV. The total answer for the suppression is then given by 
(\ref{mHcorrections}) evaluated at $\bar\chi_*$. The corrections due
to $m_{eff}$ and $H_{inf}$
are small as long
as\footnote{The current bound on the primordial tensor perturbations 
\cite{Ade:2015tva} constrains $H_{inf}\lesssim 10^{14}$~GeV during last
$\sim 60$ efolds of inflation.}  
$m_{eff}, H_{inf}\lesssim 10^{15}\div 10^{17}$~GeV.

\section{Discussion of approximations}

We have obtained the formula (\ref{mHcorrections}) under the assumption
that the transition happens in an external de Sitter space-time. Let us
check its validity. First, the Hubble rate during inflation is not
exactly constant, but slowly varies. We have seen that the
size of the bounce is much smaller that the 
horizon size. This implies that the formation of the bubble of the new
phase 
inside the false vacuum occurs very fast\footnote{The time of the
  bubble formation should not be confused with the vacuum decay time,
  which is exponentially long.}. Thus neglecting the change
in the Hubble rate during the formation of the bubble is justified. 

Second, in the case when the effective Higgs mass is given by the
coupling to the inflaton, the Higgs exerts a force on the inflaton 
during tunneling. This force should not lead to large
displacements of $\phi$ that could change its energy density. One
estimates the shift of $\phi$ due to the Higgs force as 
\be
\Box\delta\phi=\frac{h^2}{2}\frac{dm_{eff}^2}{d\phi}~~
\Longrightarrow~~
\delta\phi\sim \frac{h^2_*}{H^2}\frac{dm_{eff}^2}{d\phi}\;,
\ee  
where box stands for the Laplacian on the 4-sphere and 
$h_*=\sqrt{8/|\lambda(\bar\chi^{-1}_*)|}\,\bar\chi^{-1}_*$ 
is the value of the Higgs in the
center of the instanton. Requiring  
$V'_{inf}\delta\phi\ll V_{inf}$ we obtain the condition 
\be
\frac{dm_{eff}^2}{d\phi}\ll\frac{V'_{inf}}{6 \epsilon h_*^2}\;,
\ee
where $\epsilon=(M_pV'_{inf})^2/(16\pi V_{inf}^2)$ is the slow-roll
parameter. 
This condition is
satisfied if the dependence of $m_{eff}$ on the inflaton is weak enough. 

Last, but not least, one should check if the energy density of the
Higgs field is smaller than that of the inflaton. This requirement
turns out to be 
violated in the center of the CDL bounce for realistic values
of $H_{inf}$. What
saves the day is the fact that the size of the region where this 
violation occurs is of order $\bar\chi_*$. On the
other hand, the log-enhanced corrections in
(\ref{mHcorrections}) come from the region of order $\sim l$, which is
much larger. Thus
they are not modified by the back-reaction of the Higgs field on the
geometry. 

The effects of the back-reaction can be taken into account neglecting
completely the inflaton energy density, i.e. in the same way as in the
case of the false vacuum decay in the flat space
\cite{Isidori:2007vm,Rajantie:2016hkj,Salvio:2016mvj}. They give an additional contribution to the
bounce action\footnote{Here we assume 
  that gravity is described by Einstein's general relativity at least
  up to the scale $\bar\chi^{-1}$.},
\be
\Delta S_E^{(CDL)}=\frac{256\pi^3(1-6\xi)^2}{45(M_p\bar\chi\lambda)^2}\;.
\ee
For moderate values of $\xi$ these corrections are small as long as
$\bar\chi_*^{-1}<5\cdot 10^{16}$~GeV. 
Finally, further corrections to the bounce action can come from
Planck-suppressed higher-order operators in the Higgs action. The
analysis of these corrections is the same as in flat space-time. Note that they can be quite significant due to the fact that the size of the instanton is close to Planckian \cite{Branchina:2013jra}.

\paragraph{Acknowledgments} We thank Fedor Bezrukov, Archil Kobakhidze, Alexander Spencer-Smith and Arttu Rajantie for stimulating discussions.  S.S. is supported by the Swiss National Science Foundation, grant No.156749.


\end{document}